\documentclass[showpacs,preprintnumbers,amsmath,amssymb,prb,twocolumn]{revtex4}

\def\XXint#1#2#3{{\setbox0=\hbox{$#1{#2#3}{\int}$}
     \vcenter{\hbox{$#2#3$}}\kern-.5\wd0}}

\usepackage{graphicx}
\usepackage{subfigure}
\usepackage{bm}

\begin{document}

\title{Control of multiferroic order by magnetic field in frustrated helimagnet MnI$_2$. Theory.}

\author{O.\ I.\ Utesov$^{1,2}$}
\email{utiosov@gmail.com}
\author{A.\ V.\ Syromyatnikov$^{1,3}$}
\email{asyromyatnikov@yandex.ru}

\affiliation{$^1$National Research Center ``Kurchatov Institute'' B.P.\ Konstantinov Petersburg Nuclear Physics Institute, Gatchina 188300, Russia}
\affiliation{$^2$St. Petersburg Academic University - Nanotechnology Research and Education Centre of the Russian Academy of Sciences, 194021 St.\ Petersburg, Russia}
\affiliation{$^3$St.\ Petersburg State University, 7/9 Universitetskaya nab., St.\ Petersburg, 199034
Russia}

\date{\today}

\begin{abstract}

We provide a theoretical description of frustrated multiferroic $\rm MnI_2$ with a spiral magnetic ordering in magnetic field $\bf h$. We demonstrate that subtle interplay of exchange coupling, dipolar forces, hexagonal anisotropy, and the Zeeman energy account for the main experimental findings observed recently in this material (Kurumaji, et al., Phys.\ Rev.\ Lett.\ {\bf 106}, 167206 (2011)). We describe qualitatively the non-trivial evolution of electric polarization $\bf P$ upon $\bf h$ rotation, changing $\bf P$ direction upon $h$ increasing, and disappearance of ferroelectricity at $h>h_c$, where $h_c$ is smaller than the saturation field.

\end{abstract}


\maketitle

\section{Introduction}
\label{intro}

Multiferroics are among the most interesting and perspective materials nowadays. The possibility of cross control between electric and magnetic degrees of freedom gives rise to various highly desirable  applications of these compounds. \cite{nagaosa} The main goal is to synthesize a material with strong magnetoelectric coupling in order to control magnetization {\bf M} (electric polarization {\bf P}) by electric (magnetic) field. \cite{Khomskii} Very promising in this respect are multiferroics of spin origin in which ferroelectricity is induced by spiral magnetic ordering and a giant magnetoelectric response was discovered.
Three main mechanisms of ferroelectricity of spin ordering are discussed now: exchange-striction mechanism, the inverse Dzyaloshinskii-Moriya (DM) mechanism, and the spin-dependent p-d hybridization mechanism. \cite{nagaosa}

Frustration plays an important role in many multiferroics of spin origin. In particular, the frustration producing a short-period spiral magnetic ordering is indispensable for the inverse DM mechanism of ferroelectricity. \cite{Mostovoy}
Besides, the frustration-induced proper screw type of magnetic ordering can lead to ferroelectricity in some materials through the variation in the metal-ligand hybridization with spin-orbit coupling (the spin-dependent p-d hybridization mechanism). \cite{nagaosa,Arima2007}
Due to the high symmetry of crystal lattice, such compounds can host multiple domains with different electric polarizations. This leads to possibility of switching between the domains (i.e., changing $\bf P$ of the whole sample) by magnetic field $\bf H$. Such $\bf H$-induced rearrangement of six domains was studied experimentally \cite{Seki2009} in triangular lattice helimagnet CuFe$_{1-x}$Ga$_x$O$_2$. At large enough in-plane $\bf H$, the electric polarization $\mathbf{P} || \mathbf{q}$ flops on $120^\circ$ upon $\bf H$ rotation through $(60p)^\circ$, where $p$ is integer, as a result of a switch of the helical vector $\mathbf{q}$.

The situation is more complicated in another triangular lattice helimagnet MnI$_2$.
At $H=0$, the proper screw magnetic ground state hosts in-plane electric polarization $\mathbf{P} \perp \mathbf{q} || \langle 1 \overline{1} 0\rangle$ which can be accounted for by both the inverse DM and the p-d hybridization mechanisms. \cite{mni3,Utesov2017} Corresponding six domains can be controlled by magnetic field $H<3$~T in the manner described above. \cite{mni3} However, the stable $\bf q$ direction changes to $\langle 1 1 0\rangle$ and $\mathbf{P}$ becomes parallel to $\mathbf{q}$  when $H$ exceeds $3$~T. At $H\approx3$~T, $ \mathbf{q}$ and $\mathbf{P}$ rotate smoothly upon in-plane $\bf H$ rotation.

In the present paper, we address this peculiar competition of two multiferroic orders in MnI$_2$ in magnetic field. We use the model which we propose in Ref.~\cite{Utesov2017} for the description of the successive phase transitions in helimagnets with dipolar interaction (which successfully describes also  MnI$_2$). We perform below a ground-state energy analysis of the system taking into account symmetry-allowed anisotropic interactions (magnetic dipolar interaction as well as easy-axis and hexagonal anisotropies), which were shown to play an important role in MnI$_2$. \cite{Utesov2017} We describe quantitatively the experimentally observed low-temperature behavior of MnI$_2$ in the external magnetic field.

The rest of the paper is organized as follows. We discuss $\rm MnI_2$ in zero and finite magnetic fields in Secs.~\ref{theor} and \ref{fieldsec}, respectively. Sec.~\ref{conc} contains summary and conclusion.

\section{$\rm MnI_2$ at zero magnetic field}
\label{theor}

\begin{figure}
  \centering
  \includegraphics[width=7cm]{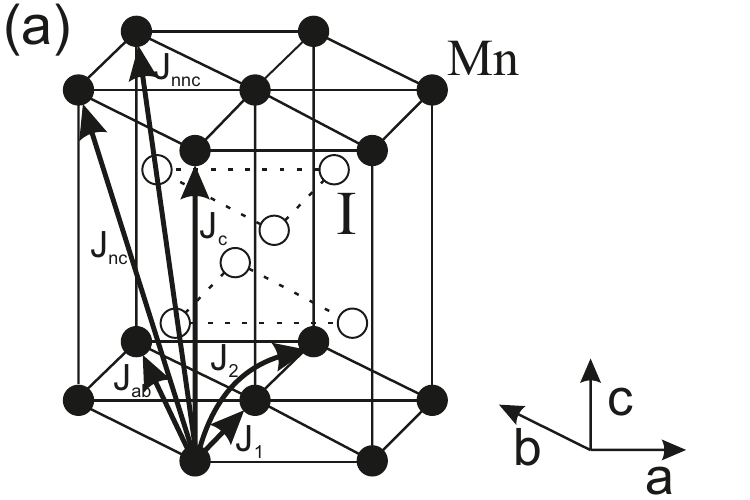}
  \includegraphics[width=6cm]{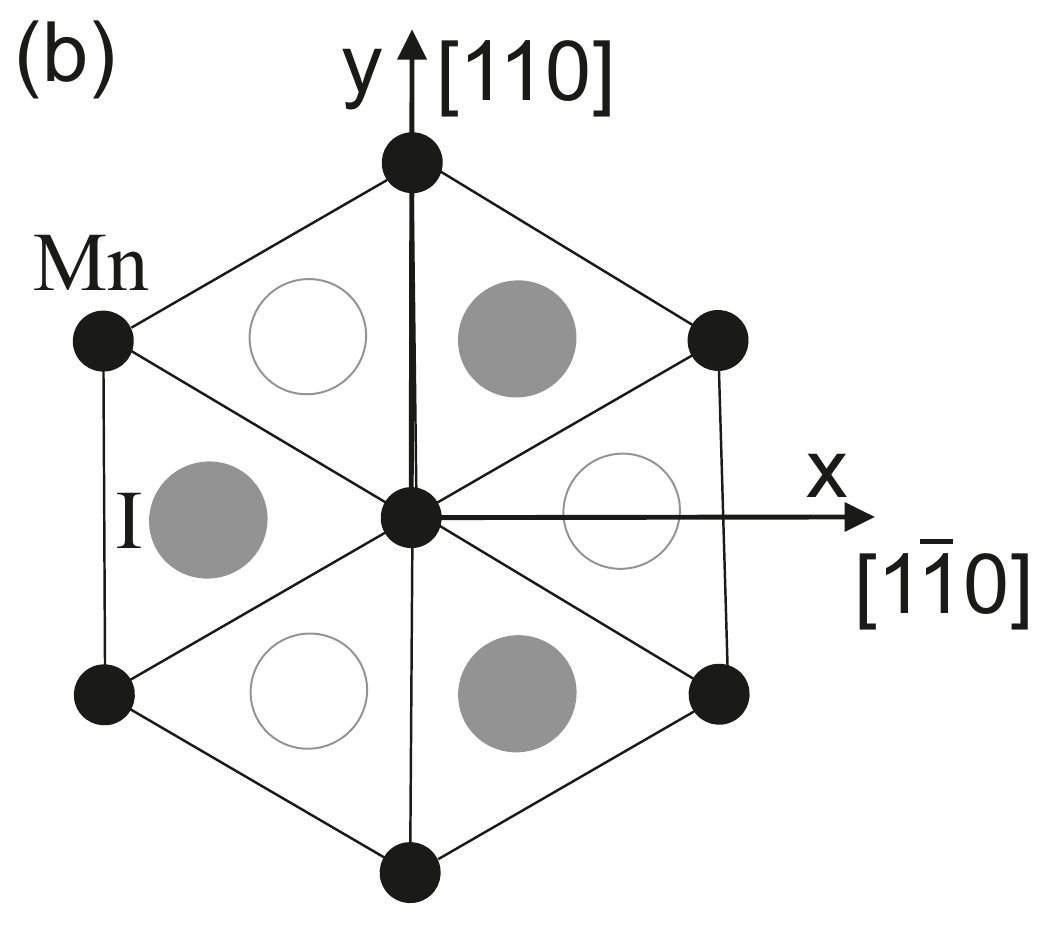}
  \caption{(a) Crystal structure of MnI$_2$. Exchange interactions $J$ are also shown. (b) Triangular layer of MnI$_2$ structure. Black circles stand for magnetic Mn$^{2+}$ ions. Gray and white circles are iodide ions located below and above the triangular plane, respectively.}
\label{FMnI2}
\end{figure}

MnI$_2$ crystallizes in a layered hexagonal lattice with centrosymmetric space group $P\overline{3}m1$ (see Fig.~\ref{FMnI2}(a)). Triangular planes of magnetic Mn$^{2+}$ ions are stacked along hexagonal $z$ axis. Positions of ligand iodide ions alternate above and below the planes as it is shown in Fig.~\ref{FMnI2}(b). Mn$^{2+}$ ions are in spherically symmetric state with $L=0$, $S=5/2$, and $g \approx 2$ that makes the spin-orbit interaction quite small. As a result, the dipole interaction becomes one of the main source of anisotropy in MnI$_2$. According to the neutron diffraction data, \cite{sato} this compound undergoes three successive magnetic phase transitions at $T_{N1}=3.95$~K, $T_{N2}=3.8$~K, and $T_{N3}=3.45$~K. The second-order transition takes place at $T=T_{N1}$ to the phase with an incommensurate sinusoidally-modulated (ICS) spin order in which the magnetization is directed along the twofold symmetry axis of the magnetic subsystem. The second-order transition at $T=T_{N2}$ is related with the breaking of the twofold rotational symmetry in the ICS phase: at $T_{N3}<T<T_{N2}$, the projection of the ICS modulation vector $\mathbf{q}$ onto $xy$-plane and the magnetization continuously move upon $T$ decreasing from one high-symmetry direction to another (e.g., from $\langle 1 0 0 \rangle$ to  $\langle 1 \overline{1} 0 \rangle$). At $T=T_{N3}$, the first-order transition occurs to a phase with a proper screw magnetic ordering in which spins rotate in the plane perpendicular to $\mathbf{q} \approx (0.181, 0, 0.439)$. The proper screw spin texture breaks the inversion symmetry, thus allowing for the electric polarization along $[110]$ axis. \cite{mni3}.

This cascade of magnetic phase transitions was successfully described theoretically within a mean-field theory in our previous paper \cite{Utesov2017}. We demonstrated that due to the small exchange integrals (which would lead only to a single transition to the spiral phase) spin interactions of relativistic nature are responsible for the set of phase transitions. The essential ingredients of our model were: (i) magnetic dipole interaction which provides correct magnetization direction in the ICS phase, (ii) in-plane hexagonal anisotropy which is responsible for the transition at $T_{N2}$ by making $[110]$ set of axes (see Fig.~\ref{FMnI2}) to be easy directions of the magnetization, (iii) easy-axis anisotropy which determines the spin rotational plane in the low-$T$ spiral phase.

We adopt this model below to describe MnI$_2$ at small $T$ in the external magnetic field $\mathbf{H}$. The corresponding Hamiltonian of the magnetic subsystem reads as
\begin{eqnarray}
\label{ham1}
  \mathcal{H} &=& \mathcal{H}_{ex} + \mathcal{H}_{dip} + \mathcal{H}_{eas} + \mathcal{H}_{hex} + \mathcal{H}_{Z}, \\
  \mathcal{H}_{ex} &=& \frac{1}{2} \sum_{i,j} J_{ij} \mathbf{S}_i \cdot \mathbf{S}_j, \\
  \mathcal{H}_{dip} &=& \frac12 \sum_{i,j} D^{\alpha \beta}_{ij} S^\alpha_i S^\beta_j, \\
  \mathcal{H}_{eas} &=& - Y \sum_i \left( S^z_i \right)^2, \\
  \mathcal{H}_{hex} &=& - Z \sum_i (S^y_i)^2 \left[ (S^y_i)^2-3(S^x_i)^2 \right]^2, \\
  \mathcal{H}_{Z} &=& - \sum_i \mathbf{h} \cdot \mathbf{S}_i,
\end{eqnarray}
where $\mathcal{H}_{ex}$ is the exchange interaction (see Fig.~\ref{FMnI2}(a) for the exchange coupling interactions included in the model), $\mathcal{H}_{dip}$ is the dipolar interaction,
\begin{equation}
  D^{\alpha \beta}_{ij} = \omega_0 \frac{v_0}{4\pi}
	\left( \frac{1}{R^3_{ij}} - \frac{3 R^\alpha_{ij} R^\beta_{ij}}{R^5_{ij}}\right),
\end{equation}
$v_0$ is the unit cell volume,
\begin{equation}
  \omega_0=4\pi\frac{(g \mu_B)^2}{v_0} \approx 0.31~\text{K},
\end{equation}
is the characteristic dipolar energy (to calculate this quantity one should take lattice parameters $a=4.146$~\AA, $c=6.829$~\AA, and $g\approx2$),  $\mathcal{H}_{eas}$ and $\mathcal{H}_{hex}$ describe the easy-axis and the hexagonal anisotropies, respectively, and $\mathbf{h}=g \mu_B \mathbf{H}$ in the Zeeman energy. We omit in Eq.~\eqref{ham1} a small DM spin interaction which does not effect the spin textures and which arises at $T<T_{N3}$ as a result of the electric polarization stabilization via the inverse DM mechanism. \cite{Utesov2017} After the Fourier transform
\begin{equation}\label{four1}
  \mathbf{S}_i=\frac{1}{\sqrt{N}} \sum_\mathbf{q} \mathbf{S}_\mathbf{q} e^{i\mathbf{q} \mathbf{R}_i}
\end{equation}
contribution to the the classical energy $\mathcal{E}$ from the first three terms in Eq.~\eqref{ham1} acquires the form
\begin{equation}
\label{ham2}
  \mathcal{E}_0 = \sum_\mathbf{q} \mathcal{H}^{\alpha \beta}_\mathbf{q} S^\alpha_\mathbf{q} S^\beta_\mathbf{-q},
\end{equation}
where symmetrical tensor
$
\mathcal{H}^{\alpha \beta}_\mathbf{q}
=
\frac12 (J_\mathbf{q} \delta_{\alpha \beta}
+
D^{\alpha \beta}_\mathbf{q})
-
Y \delta_{\alpha z}  \delta_{\beta z}
$
possesses three eigenvalues $\lambda_{1,2,3}(\mathbf{q})$ and the corresponding eigenvectors $\mathbf{v}_{1,2,3}({\mathbf{q}})$, where $J_\mathbf{q} = \sum_{j\neq0} J_{0j} e^{i \mathbf{q}\mathbf{R}_j}$ and
$
  D^{\alpha \beta}_\mathbf{q} = \sum_{j\neq0} D^{\alpha \beta}_{0j} e^{i \mathbf{q}\mathbf{R}_j}
$. Slowly convergent lattice sums in the dipolar term have been rewritten in fast convergent forms (see, e.g., Ref.~\cite{cohen}) and calculated numerically. We assume below that the smallest and the largest eigenvalues are $\lambda_1(\mathbf{q})$ and  $\lambda_3(\mathbf{q})$, respectively.

By minimizing the mean-field free energy, we found in Ref.~\cite{Utesov2017} a set of parameters of the model \eqref{ham1} which successfully describes the cascade of phase transitions in MnI$_2$ at $h=0$. We found also that a lot of distinct set of parameters can be suggested to describe by model \eqref{ham1} the proper screw spiral ordering with ${\bf q}={\bf q}_{sp}=(0.181,0,0.439)$ observed experimentally at small $T$. We quoted in Ref.~\cite{Utesov2017} (see Eq.~(36)) that set of parameters which is closest to the parameters suggested for description of the high-temperature behavior of MnI$_2$.
\footnote{Notice that all $J$ values in Eqs.~(34) and (36) of Ref.~\cite{Utesov2017} should be twice as large. This difference is related to the erroneously omitted factor of 1/2 before $J_{\bf q}$ in Eq.~(5) of Ref.~\cite{Utesov2017}. All other conclusions of Ref.~\cite{Utesov2017} are not affected by this omission.}
However, as we find in the present study, that set of parameters underestimates significantly critical magnetic fields of MnI$_2$ at low temperatures. We demonstrate below that the following parameters describe successfully low-$T$ experimental findings:
\begin{equation}
\label{param2}
\begin{aligned}
  J_1&=-0.38, \quad J_2=0.42, \quad J_{ab}=-0.06, \\
  J_c&=0.28, \quad J_{nc}=0.08, \quad J_{nnc}=0.12, \\
  Y&=0.12, \quad Z=0.02,
\end{aligned}
\end{equation}
where all values are in Kelvins.

The set of different exchange couplings is illustrated in Fig.~\ref{FMnI2}(a). The most long-range exchange integral $J_{nnc}$ (which could be omitted at the first glance) plays an important role: it lowers the symmetry of the exchange interaction around the $c$-axis to the three-fold one. \cite{mnbr2} The easy-axis anisotropy determines the spiral rotational plane: if $Y$ was zero, spins would lie in $xy$-plane. $Z$ looks small but the contribution to the system energy from the six-fold anisotropy is noticeable because it is proportional to $S^6$ and $S=5/2$.




\section{$\rm MnI_2$ in magnetic field at small $T$}
\label{fieldsec}

Only perpendicular to the spin rotational plane component of the magnetic field
\begin{equation}
\label{mag1}
  \mathbf{h}=h (\cos{\phi_h},\sin{\phi_h},0)
\end{equation}
makes the main contribution to the classical energy at small enough $h$ (we consider below $h$ much smaller than the saturation field $h_s$ which is larger than 6~T in $\rm MnI_2$).
Then, the short-period spin texture is approximately conical which can be characterized by the cone angle $\alpha$ ($\alpha=0$ at $h=0$) and vector
\begin{equation}
\label{norm}
  \mathbf{n} = (\sin{\theta} \cos{\varphi} , \sin{\theta} \sin{\varphi} , \cos{\theta} )
\end{equation}
normal to the spin rotational plane ($\theta \approx 38^\circ $ and $\varphi=(60p)^\circ$ at $h=0$, where $p$ is integer). Numerical calculation with parameters \eqref{param2} shows that dipolar forces provide $\mathbf{v}_1({\bf q}) \perp {\bf q}$ so that
\begin{equation}
\label{q|n}
	{\bf q}\|\pm{\bf n}
\end{equation}
(spirals with modulation vectors ${\bf q}$ and $-{\bf q}$ have the same energy).
Then, magnetic moment at $i$-th site has the form
\begin{equation}\label{spin1}
  \mathbf{S}_i = S\left[ \left( \mathbf{a} \sin \mathbf{q}\mathbf{R}_i + \mathbf{b} \cos \mathbf{q}\mathbf{R}_i \right) \cos{\alpha} + \mathbf{n} \sin{\alpha} \right],
\end{equation}
where
\begin{equation}
\begin{array}{ll}
  \mathbf{a} &= (-\sin{\varphi}, \cos{\varphi},0) ,\\
  \mathbf{b} &= ( - \cos{\theta} \cos{\varphi} , - \cos{\theta} \sin{\varphi} , \sin{\theta} )
\end{array}
\end{equation}
are basis vectors in the spin rotation plane.

\begin{figure}
  \centering
  \includegraphics[width=8cm]{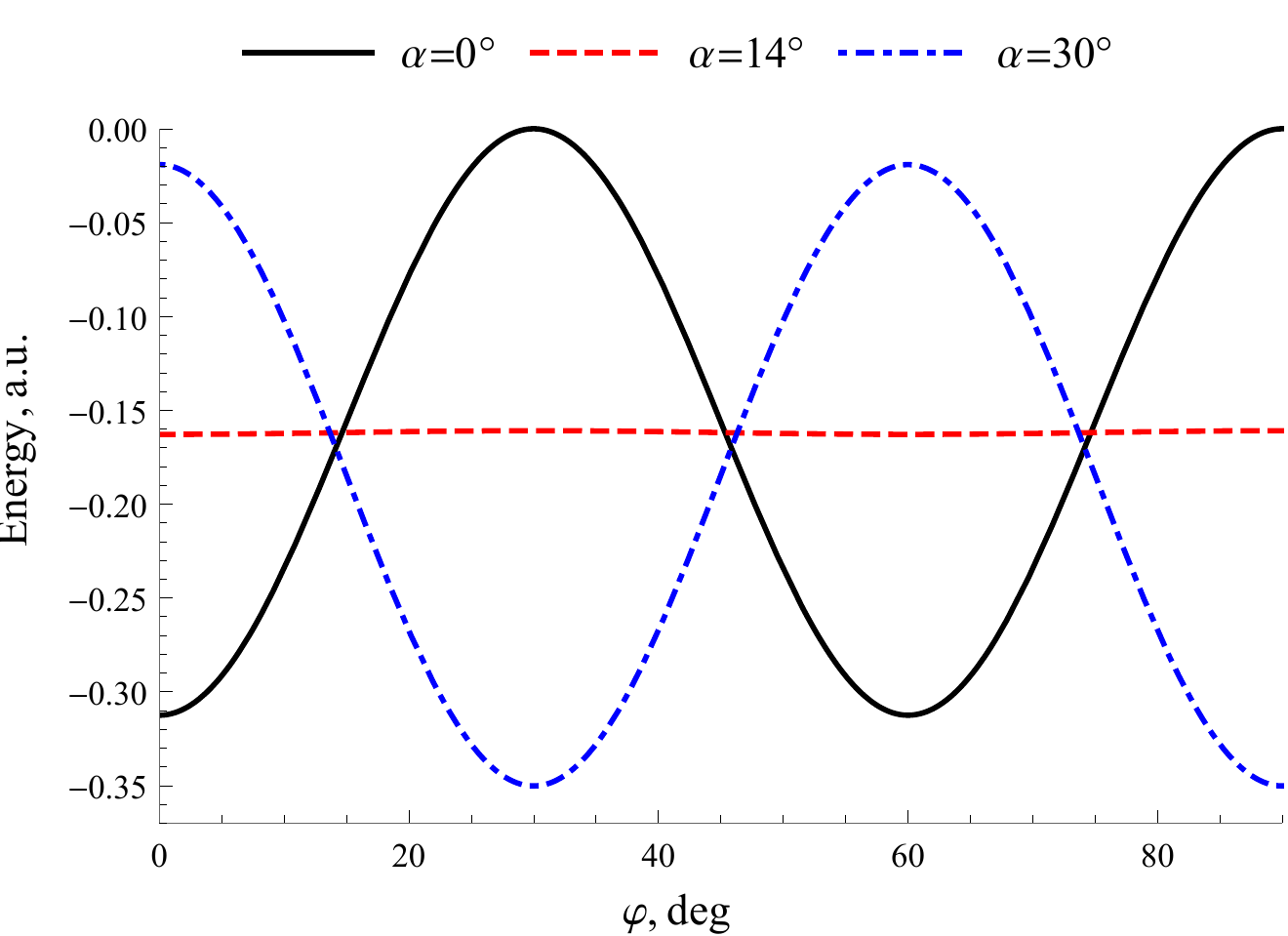}\\
  \caption{Graphics of $f(\theta, \varphi, \alpha)$ at $\theta = 38^\circ$ which appears in classical energy \eqref{meanen} and describes the hexagonal anisotropy. Under magnetic field (i.e., the cone angle $\alpha$) increasing, minima and maxima of $f$ change places so that $f$ goes through a flat profile at $\alpha\sim\alpha_c\approx15^\circ$.}
	\label{FAnis}
\end{figure}

It is seen from Eq.~\eqref{spin1} that the classical energy depends only on $\mathbf{S}_{\pm \mathbf{q}}$ and $\mathbf{S}_{\bf 0}$ and its explicit form is
\begin{eqnarray}
\label{meanen}
  && \mathcal{E}(\theta,\varphi,\alpha,{\bf h}) =
	\nonumber\\
	&&{}
	\frac{S^2}{2} \left( \lambda_1({\bf q}) + \lambda_2({\bf q}) \sin^2{\theta} + \lambda_3({\bf q}) \cos^2{\theta} \right) \cos^2{\alpha}
	\nonumber\\
	&&{}+ S^2 \lambda_0 \sin^2 \alpha - S^2 \left(Y - \frac{{\cal N}_{zz} \omega_0}{2} \right) \cos^2 \theta \sin^2 \alpha \nonumber  \\
	&&{} -Z S^6 f(\theta, \varphi, \alpha)-h S \sin{\theta} \sin{\alpha} \cos{\left(\varphi - \phi_h \right)}  \\
	&&{}- X S^4 \sin^2{2 \theta}, \nonumber
\end{eqnarray}
where ${\cal N}_{zz}$ is the demagnetization tensor component, the term with $f(\theta, \varphi, \alpha)$ originates from the hexagonal anisotropy (see Fig.~\ref{FAnis} for graphics of this function at $\theta= 38^\circ $), and $\lambda_0 = J_{\bf 0} - \omega_0/3 $. For definiteness, we consider below a crystal having the form of a thin plate so that ${\cal N}_{zz}=1$.

The last term in Eq.~\eqref{meanen} describes phenomenologically the contribution from the magnetoelectric coupling. It is derived as follows. In general, electric polarization can be found by minimization of the mean energy \cite{Mostovoy}
\begin{equation}\label{MEcoupl}
  \mathcal{H}_{ME} =  - \gamma \mathbf{P} \cdot \mathbf{P}_{SO} + \frac{P^2}{2 \chi_E},
\end{equation}
where the first term arises due to the non-collinear spin ordering. For ligand lying between $i$-th and $j$-th magnetic ions, the inverse DM mechanism gives \cite{nagaosa}
\begin{equation}
\label{psodm}
	\mathbf{P}_{SO} \propto \mathbf{e}_{ij} \times \left[ \mathbf{S}_i \times \mathbf{S}_j \right]
\end{equation}
whereas the p-d hybridization mechanism leads to \cite{nagaosa}
\begin{equation}
\label{psopd}
	\mathbf{P}_{SO} \propto (\mathbf{e}_{il} \cdot \mathbf{S}_i)^2 \mathbf{e}_{il},
\end{equation}
where $l$ denotes the site with nonmagnetic ligand. One obtains by minimization of Eq.~\eqref{MEcoupl} $\mathbf{P}=\gamma \chi_E\mathbf{P}_{SO}$ and  $\mathcal{H}_{ME}=-\chi_E\gamma^2P^2_{SO}/2$. Eqs.~\eqref{psodm} and \eqref{psopd} provide different $\theta$-dependence of $\mathbf{P}_{SO}$ but it follows from the symmetry that both Eqs.~\eqref{psodm} and \eqref{psopd} vanish for spirals with $\theta=\pi/2$ and $\theta=0$ (in particular, it is shown in Ref.~\cite{Utesov2017} that the inverse DM mechanism gives $P \propto \sin \left(\frac{\sqrt{3}}{2} q\sin\theta \right) \cos\theta$ which vanishes at $\theta=0,\pi/2$ as it is seen from Eqs.~\eqref{norm} and \eqref{q|n}). The simplest function which describes phenomenologically this $\theta$-dependence is $\sin{2 \theta}$ (calculations show that our conclusions are insensitive to the particular choice of this function). Then, it is seen from Eqs.~\eqref{psodm} and \eqref{psopd} that $P_{SO}\propto S^2$. As a result, we come from $\mathcal{H}_{ME}=-\chi_E \gamma^2 P^2_{SO}/2$ to the last term in Eq.~\eqref{meanen}, where $X$ is a constant.

The following qualitative picture arises from analysis of Eq.~\eqref{meanen}. At small $h$, $\alpha$ is also small and the hexagonal anisotropy provides energy minima at $\varphi = (60p)^\circ$, where $p$ is integer, that corresponds to $\mathbf{q}^\perp || \langle 1 \overline{1} 0\rangle$ directions (see Figs.~\ref{FMnI2} and \ref{FAnis}), where $\mathbf{q}^\perp$ is the projection of ${\bf q}$ on the triangular plane. Then, six domains can appear in a sample with six possible orientations of $\mathbf{q}$. Electric polarization $\mathbf{P} \perp \mathbf{q}^\perp$ arises in each domain via the inverse DM or the p-d hybridization mechanisms. \cite{nagaosa,Utesov2017} The energy maxima are at $\mathbf{q}^\perp || \langle 1 1 0\rangle$ so that switches of $\mathbf{q}$ take place upon magnetic field rotation across angles $\phi_h=(30+60p)^\circ$ which are accompanied with switches of $\bf P$ (it is seen from Eq.~\eqref{meanen} that the Zeeman term is minimized upon such $\mathbf{q}$ flops).

Naively, one could expect that upon a $60^\circ$ counterclockwise rotation of $\bf h$, $\bf P$ turns by the same angle of $60^\circ$ in the same direction. However, the situation is more complicated. Because the system possesses only a three-fold rotation axis (see Fig.~\ref{FMnI2}), the $60^\circ$ rotation of the magnetic field could be accompanied with the same rotation of $\mathbf{q}^\perp$ but $q_z$ should change its sign to provide the energy minimum. Such evolution of $\bf q$, however, would be at odds with the experiment which shows that spin chirality is preserved upon $\bf h$-induced $\bf q$-flops due to the difference in stability between two different multiferroic domain walls~\cite{Seki2009,mni3}. The evolution of $\bf q$ which provides minima for both domain energy and the domain walls energy is changing the sign of $\mathbf{q}^\perp$ and preserving $q_z$. It is easy to show that such $\bf q$ modification leads to a clockwise rotation by $120^\circ$ of $\bf P$.

\begin{figure}
  \centering
  \includegraphics[width=8cm]{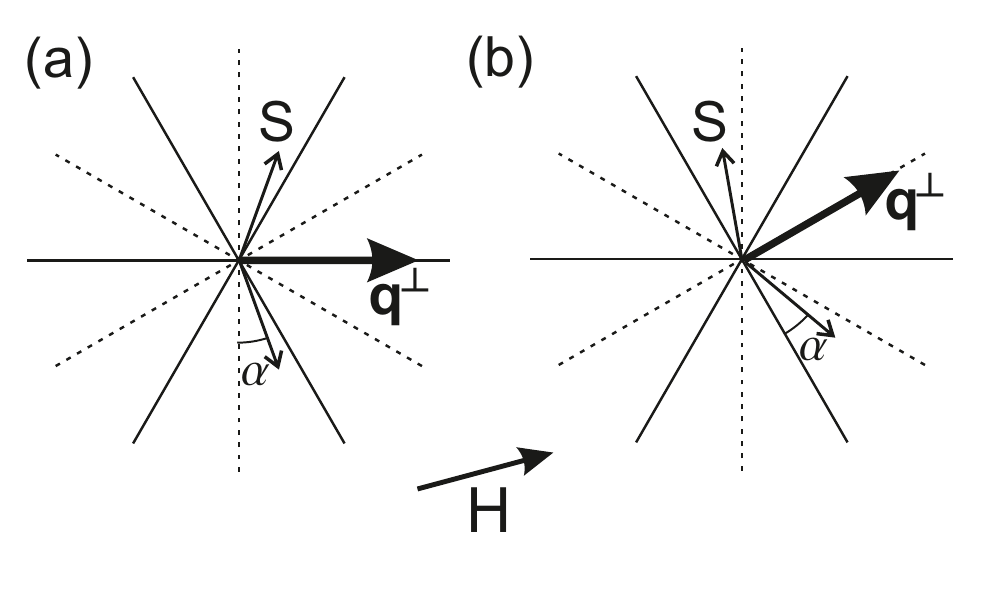}\\
  \caption{Triangular planes are shown, where dashed and solid lines are easy and hard directions, respectively, made by the hexagonal anisotropy. This figure illustrates the changing places of minima and maxima of the hexagonal anisotropy energy upon field increasing which is shown in Fig.~\ref{FAnis} (see the text).}
	\label{FCones}
\end{figure}

Under $h$ increasing, $\alpha$ also rises and minima and maxima of the hexagonal anisotropy change places at some critical value $\alpha_c \approx 15^\circ$ of the cone angle as it is seen from Fig.~\ref{FAnis}. This change of the hexagonal anisotropy can be explained qualitatively by simple example of the proper screw spiral with $\theta=90^\circ$ (see Fig.~\ref{FCones}). Projections of spins on the triangular plane are shown in Fig.~\ref{FCones} for two cases: (a) ${\bf h}\|{\bf q}^\perp\|\langle 1 \bar1 0\rangle$ (this configuration minimizes the anisotropy energy at $h=0$); (b) ${\bf h}\|{\bf q}^\perp\|\langle 1 1 0\rangle$ (this configuration maximizes the anisotropy energy at $h=0$). It is seen from Fig.~\ref{FCones} that upon field increasing, spins in the cone spiral become closer to the hard and easy directions in cases (a) and (b), respectively, that results in the changing places of minima and maxima at the critical cone angle $\alpha_c$.

\begin{figure}
  \centering
  \includegraphics[width=8cm]{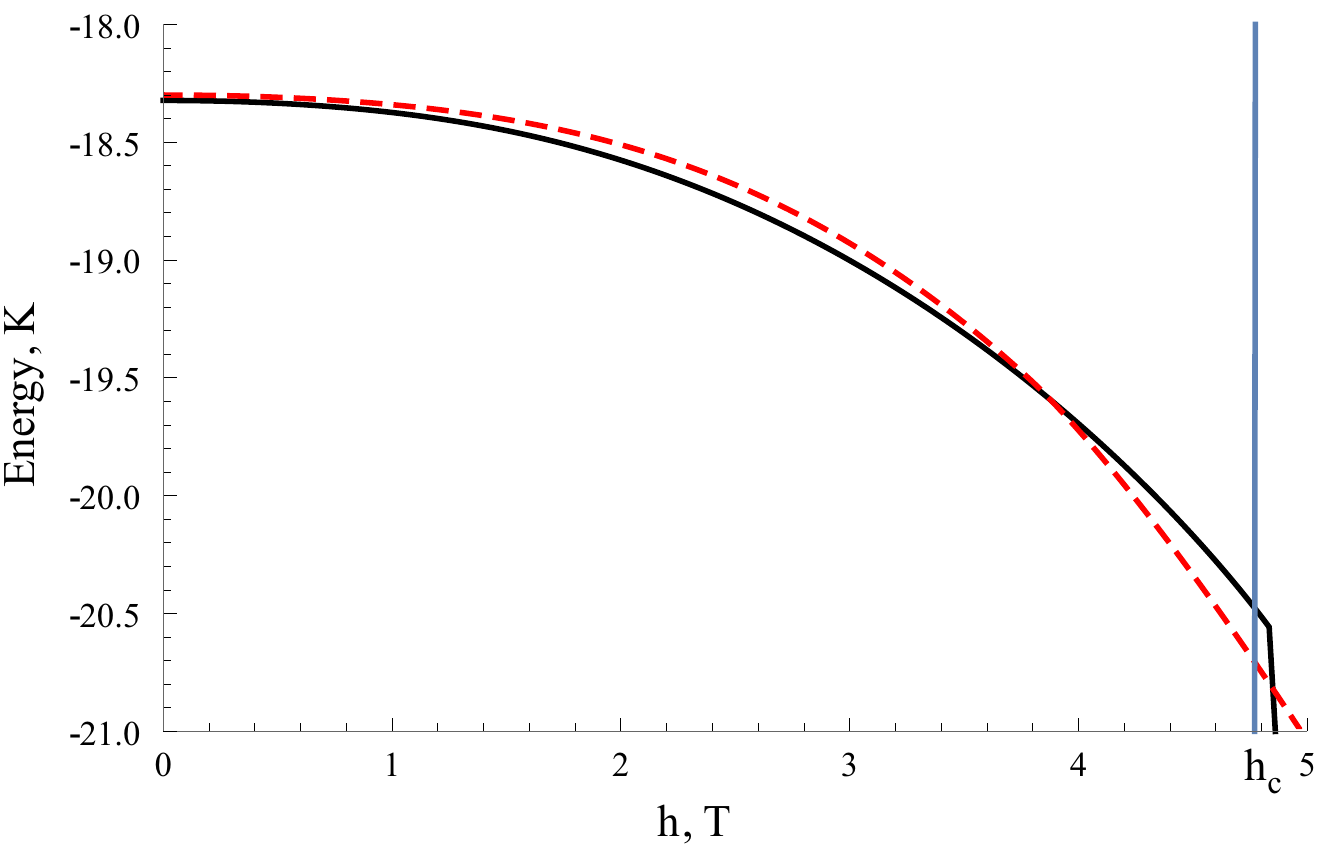}\\
  \caption{Classical energies of two spin states with ${\bf q}^\perp\|\langle 1 \bar1 0\rangle$ and ${\bf q}^\perp\|\langle 1 1 0\rangle$ having different multiferroic orderings ($\mathbf{P} \perp \mathbf{q}^\perp$ and $\mathbf{P} || \mathbf{q}^\perp$, respectively) at $\mathbf{h} \| \langle 1 \overline{1} 0\rangle$. At small magnetic field, configuration with $\mathbf{q}^\perp || \langle 1 \overline{1} 0\rangle $ (black curve) is stable, whereas at $h \approx 3.8$~T a first-order transition takes place to the state with $\mathbf{q}^\perp\|\langle 1 1 0\rangle$ (red curve) due to subtle interplay between Zeeman energy and the hexagonal anisotropy. At $h=h_c \approx 4.8$~T, $\theta$ becomes equal to $90^\circ$ in both spin textures in which case $\mathbf{P}=0$.}
	\label{FPhase}
\end{figure}

\begin{figure}
  \centering
  \includegraphics[width=4cm]{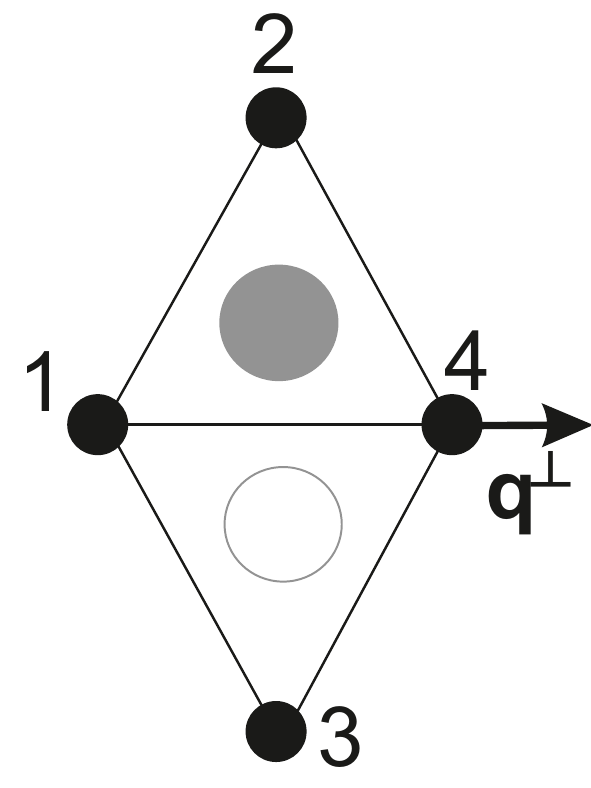}\\
  \caption{Cluster for derivation of $\bf P$ when $\mathbf{q}^\perp || \langle 1 1 0\rangle$.}
	\label{FPDhyb}
\end{figure}

The system classical energy changes accordingly upon $h$ increasing: at moderate magnetic fields $h\agt2.6$~T, the system energy becomes smaller in configuration ${\bf h}\|{\bf q}^\perp\|\langle 1 1 0\rangle$ than in configuration ${\bf h}\|{\bf q}^\perp\|\langle 1 \bar1 0\rangle$. As a result, at a given orientation of $\bf h$ spin ordering changes in domains at some field value from ${\bf q}^\perp\|\langle 1 \bar1 0\rangle$ to ${\bf q}^\perp\|\langle 1 1 0\rangle$ (see Fig.~\ref{FPhase}).  The in-plane field rotation leads to the switches between these domains at $\phi_h = (60p)^\circ$. This changing of the spin ordering is accompanied with changing the type of the multiferroic ordering in the domains from $\mathbf{P} \perp \mathbf{q}^\perp$ to $\mathbf{P} || \mathbf{q}^\perp$. The latter can be described by the p-d hybridization mechanism (see, e.g., Ref.~\cite{Arima2007}). It is seen from Fig.~\ref{FMnI2} that one has to consider the cluster shown in Fig.~\ref{FPDhyb} to determine $\bf P$ direction at $\mathbf{q}^\perp\|\langle 1 1 0\rangle$. It is easy to show using Eq.~\eqref{psopd} that contributions to the $\bf P$ component perpendicular to $\mathbf{q}^\perp$ exactly cancel each other, and only $\bf P$ component along $\mathbf{q}^\perp$ can be nonzero. The clockwise rotation by $120^\circ$ of $\bf P$ upon the counterclockwise rotation of $\bf h$ by $60^\circ$ is explained in the same spirit as it is done above for the low-field regime.

At $h\sim 2.6$~T both spin textures have approximately the same energy and the hexagonal anisotropy is almost independent of $\varphi$ (see Figs.~\ref{FAnis} and \ref{FPhase}). This results in a continuous rotation of $\mathbf{q}^\perp$ by rotating magnetic field keeping $\mathbf{q}^\perp \| \mathbf{h}$. As it is explained above, $\bf P\perp\mathbf{q}^\perp$ and $\bf P\|\mathbf{q}^\perp$ when ${\bf q}^\perp\|\langle 1 \bar1 0\rangle$ and ${\bf q}^\perp\|\langle 1 1 0\rangle$, respectively. Then, $\bf P$ rotates clockwise twice while magnetic field rotates counterclockwise once at this critical field region. This picture is in full qualitative agreement with experimental findings of Ref.~\cite{mni3}.

Under magnetic field increasing, $\theta$ also grows so that vector $\mathbf{n}$ lays on the $xy$-plane ($\theta=90^\circ$) at large enough $h$. In this case, electric polarization is exactly zero in accordance with both the inverse DM and the p-d hybridization mechanisms. Then, the field value $h_c$ (denoted in Fig.~\ref{FPhase}) at which $\theta$ becomes equal to $90^\circ$ determines the border of the ferroelectric phase.

According to our calculations with parameters \eqref{param2}, $h_c\approx 4.8$~T whereas its experimentally obtained \cite{mni3} value is slightly smaller than 6~T. Theoretically obtained field value of $\approx2.6$~T at which the smooth rotation of $\mathbf{q}^\perp$ occurs is close to the corresponding experimental value of $\approx3$~T. We observe the first-order transition to the state with $\mathbf{q}^\perp\|\langle 1 1 0\rangle$ at $h\approx3.8$~T (see Fig.~\ref{FPhase}) that is also in good agreement with the experiment (according to Ref.~\cite{mni3}, this field lies in the interval $(2.7,3.3)$~T).
Overall, the quantitative consistency of our discussion with the experiment is reasonably good.


\section{Summary and conclusion}
\label{conc}

To conclude, we provide a theoretical description of multiferroic $\rm MnI_2$ in magnetic field $\bf h$ at small $T$. We show that a subtle interplay of exchange coupling, dipolar forces, hexagonal anisotropy, and the Zeeman energy account for key experimental findings of Ref.~\cite{mni3}. In particular, we show that $\bf P$ turns by $120^\circ$ clockwise upon the counterclockwise field rotation by $60^\circ$. We demonstrate that $\bf P$ direction changes upon $h$ increasing from $\mathbf{P} \perp \mathbf{q}^\perp$ to $\mathbf{P} || \mathbf{q}^\perp$. It is also observed that the ferroelectricity disappears at $h>h_c$, where $h_c$ is smaller than the saturation field.

\begin{acknowledgments}

The reported study was funded by RFBR according to the research project  18-02-00706.

\end{acknowledgments}

\appendix

\bibliography{TAFbib}

\begin{thebibliography}{10}
\expandafter\ifx\csname natexlab\endcsname\relax\def\natexlab#1{#1}\fi
\expandafter\ifx\csname bibnamefont\endcsname\relax
  \def\bibnamefont#1{#1}\fi
\expandafter\ifx\csname bibfnamefont\endcsname\relax
  \def\bibfnamefont#1{#1}\fi
\expandafter\ifx\csname citenamefont\endcsname\relax
  \def\citenamefont#1{#1}\fi
\expandafter\ifx\csname url\endcsname\relax
  \def\url#1{\texttt{#1}}\fi
\expandafter\ifx\csname urlprefix\endcsname\relax\def\urlprefix{URL }\fi
\providecommand{\bibinfo}[2]{#2}
\providecommand{\eprint}[2][]{\url{#2}}

\bibitem[{\citenamefont{Tokura et~al.}(2014)\citenamefont{Tokura, Seki, and
  Nagaosa}}]{nagaosa}
\bibinfo{author}{\bibfnamefont{Y.}~\bibnamefont{Tokura}},
  \bibinfo{author}{\bibfnamefont{S.}~\bibnamefont{Seki}}, \bibnamefont{and}
  \bibinfo{author}{\bibfnamefont{N.}~\bibnamefont{Nagaosa}},
  \bibinfo{journal}{Reports on Progress in Physics}
  \textbf{\bibinfo{volume}{77}}, \bibinfo{pages}{076501}
  (\bibinfo{year}{2014}), \bibinfo{note}{and references therein}.

\bibitem[{\citenamefont{Khomskii}(2006)}]{Khomskii}
\bibinfo{author}{\bibfnamefont{D.}~\bibnamefont{Khomskii}},
  \bibinfo{journal}{Journal of Magnetism and Magnetic Materials}
  \textbf{\bibinfo{volume}{306}}, \bibinfo{pages}{1 } (\bibinfo{year}{2006}),
  ISSN \bibinfo{issn}{0304-8853},
  \urlprefix\url{http://www.sciencedirect.com/science/article/pii/S0304885306004239}.

\bibitem[{\citenamefont{Mostovoy}(2006)}]{Mostovoy}
\bibinfo{author}{\bibfnamefont{M.}~\bibnamefont{Mostovoy}},
  \bibinfo{journal}{Phys. Rev. Lett.} \textbf{\bibinfo{volume}{96}},
  \bibinfo{pages}{067601} (\bibinfo{year}{2006}),
  \urlprefix\url{https://link.aps.org/doi/10.1103/PhysRevLett.96.067601}.

\bibitem[{\citenamefont{Arima}(2007)}]{Arima2007}
\bibinfo{author}{\bibfnamefont{T.}~\bibnamefont{Arima}},
  \bibinfo{journal}{Journal of the Physical Society of Japan}
  \textbf{\bibinfo{volume}{76}}, \bibinfo{pages}{073702}
  (\bibinfo{year}{2007}),
  \urlprefix\url{https://doi.org/10.1143/JPSJ.76.073702}.

\bibitem[{\citenamefont{Seki et~al.}(2009)\citenamefont{Seki, Murakawa, Onose,
  and Tokura}}]{Seki2009}
\bibinfo{author}{\bibfnamefont{S.}~\bibnamefont{Seki}},
  \bibinfo{author}{\bibfnamefont{H.}~\bibnamefont{Murakawa}},
  \bibinfo{author}{\bibfnamefont{Y.}~\bibnamefont{Onose}}, \bibnamefont{and}
  \bibinfo{author}{\bibfnamefont{Y.}~\bibnamefont{Tokura}},
  \bibinfo{journal}{Phys. Rev. Lett.} \textbf{\bibinfo{volume}{103}},
  \bibinfo{pages}{237601} (\bibinfo{year}{2009}),
  \urlprefix\url{https://link.aps.org/doi/10.1103/PhysRevLett.103.237601}.

\bibitem[{\citenamefont{Kurumaji et~al.}(2011)\citenamefont{Kurumaji, Seki,
  Ishiwata, Murakawa, Tokunaga, Kaneko, and Tokura}}]{mni3}
\bibinfo{author}{\bibfnamefont{T.}~\bibnamefont{Kurumaji}},
  \bibinfo{author}{\bibfnamefont{S.}~\bibnamefont{Seki}},
  \bibinfo{author}{\bibfnamefont{S.}~\bibnamefont{Ishiwata}},
  \bibinfo{author}{\bibfnamefont{H.}~\bibnamefont{Murakawa}},
  \bibinfo{author}{\bibfnamefont{Y.}~\bibnamefont{Tokunaga}},
  \bibinfo{author}{\bibfnamefont{Y.}~\bibnamefont{Kaneko}}, \bibnamefont{and}
  \bibinfo{author}{\bibfnamefont{Y.}~\bibnamefont{Tokura}},
  \bibinfo{journal}{Phys. Rev. Lett.} \textbf{\bibinfo{volume}{106}},
  \bibinfo{pages}{167206} (\bibinfo{year}{2011}).

\bibitem[{\citenamefont{Utesov and Syromyatnikov}(2017)}]{Utesov2017}
\bibinfo{author}{\bibfnamefont{O.~I.} \bibnamefont{Utesov}} \bibnamefont{and}
  \bibinfo{author}{\bibfnamefont{A.~V.} \bibnamefont{Syromyatnikov}},
  \bibinfo{journal}{Phys. Rev. B} \textbf{\bibinfo{volume}{95}},
  \bibinfo{pages}{214420} (\bibinfo{year}{2017}),
  \urlprefix\url{https://link.aps.org/doi/10.1103/PhysRevB.95.214420}.

\bibitem[{\citenamefont{Sato et~al.}(1995)\citenamefont{Sato, Kadowaki, and
  Iio}}]{sato}
\bibinfo{author}{\bibfnamefont{T.}~\bibnamefont{Sato}},
  \bibinfo{author}{\bibfnamefont{H.}~\bibnamefont{Kadowaki}}, \bibnamefont{and}
  \bibinfo{author}{\bibfnamefont{K.}~\bibnamefont{Iio}},
  \bibinfo{journal}{Physica B: Condensed Matter}
  \textbf{\bibinfo{volume}{213}}, \bibinfo{pages}{224 } (\bibinfo{year}{1995}).

\bibitem[{\citenamefont{Cohen and Keffer}(1955)}]{cohen}
\bibinfo{author}{\bibfnamefont{M.~H.} \bibnamefont{Cohen}} \bibnamefont{and}
  \bibinfo{author}{\bibfnamefont{F.}~\bibnamefont{Keffer}},
  \bibinfo{journal}{Phys. Rev.} \textbf{\bibinfo{volume}{99}},
  \bibinfo{pages}{1128} (\bibinfo{year}{1955}), \bibinfo{note}{and references
  therein}.

\bibitem[{\citenamefont{Sato et~al.}(1994)\citenamefont{Sato, Kadowaki, Masudo,
  and Iio}}]{mnbr2}
\bibinfo{author}{\bibfnamefont{T.}~\bibnamefont{Sato}},
  \bibinfo{author}{\bibfnamefont{H.}~\bibnamefont{Kadowaki}},
  \bibinfo{author}{\bibfnamefont{H.}~\bibnamefont{Masudo}}, \bibnamefont{and}
  \bibinfo{author}{\bibfnamefont{K.}~\bibnamefont{Iio}}, \bibinfo{journal}{J.
  Phys. Soc. Japan} \textbf{\bibinfo{volume}{63}}, \bibinfo{pages}{4583}
  (\bibinfo{year}{1994}).

\end{thebibliography}

\end{document}